\documentclass[twocolumn,aps,amsmath,amssymb,floatfix,superscriptaddress,prb,longbibliography]{revtex4-1}

\usepackage{graphicx}
\usepackage{dcolumn}
\usepackage{bm}
\usepackage{xspace}
\usepackage{hyperref}
\usepackage{color}
\usepackage{booktabs}
\usepackage{array}
\usepackage{multirow}

\def\fmo{Fe$_{2}$Mo$_{3}$O$_{8}$\xspace}
\def\fzmo{Fe$_{1.75}$Zn$_{0.25}$Mo$_{3}$O$_{8}$\xspace}

\begin{document}

\title{Magnetic Signature of Chiral Phonons Revealed by Neutron Spectroscopy in Ferrimagnetic Fe$_{1.75}$Zn$_{0.25}$Mo$_{3}$O$_{8}$}
\author{Song~Bao}
\altaffiliation{These authors contributed equally to this work.}
\email{songbao@nju.edu.cn}
\affiliation{National Laboratory of Solid State Microstructures and Department of Physics, Nanjing University, Nanjing 210093, China}
\affiliation{Collaborative Innovation Center of Advanced Microstructures and Jiangsu Physical Science Research Center, Nanjing University, Nanjing 210093, China}
\author{Junbo~Liao}
\altaffiliation{These authors contributed equally to this work.}
\affiliation{National Laboratory of Solid State Microstructures and Department of Physics, Nanjing University, Nanjing 210093, China}
\author{Zhentao~Huang}
\altaffiliation{These authors contributed equally to this work.}
\affiliation{National Laboratory of Solid State Microstructures and Department of Physics, Nanjing University, Nanjing 210093, China}
\affiliation{Institute of High Energy Physics, Chinese Academy of Sciences (CAS), Beijing 100049, China}
\affiliation{Spallation Neutron Source Science Center, Dongguan 523803, China}
\author{Yanyan~Shangguan}
\affiliation{National Laboratory of Solid State Microstructures and Department of Physics, Nanjing University, Nanjing 210093, China}
\author{Zhen~Ma}
\affiliation{Hubei Key Laboratory of Photoelectric Materials and Devices, School of Materials Science and Engineering, Hubei Normal University, Huangshi 435002, China}
\author{Bo~Zhang}
\author{Shufan~Cheng}
\author{Hao~Xu}
\author{Zihang~Song}
\author{Shuai~Dong}
\author{Maofeng~Wu}
\affiliation{National Laboratory of Solid State Microstructures and Department of Physics, Nanjing University, Nanjing 210093, China}
\author{Ryoichi~Kajimoto}
\author{Mitsutaka~Nakamura}
\affiliation{J-PARC Center, Japan Atomic Energy Agency (JAEA), Tokai, Ibaraki 319-1195, Japan}
\author{Tom~Fennell}
\affiliation{PSI Center for Neutron and Muon Sciences, Paul Scherrer Institut, 5232 Villigen PSI, Switzerland}
\author{Dmitry~Khalyavin}
\affiliation{ISIS Facility, STFC Rutherford Appleton Laboratory, Chilton, Didcot, Oxfordshire OX11 0QX, United Kingdom}
\author{Jinsheng~Wen}
\email{jwen@nju.edu.cn}
\affiliation{National Laboratory of Solid State Microstructures and Department of Physics, Nanjing University, Nanjing 210093, China}
\affiliation{Collaborative Innovation Center of Advanced Microstructures and Jiangsu Physical Science Research Center, Nanjing University, Nanjing 210093, China}

\begin{abstract}
\noindent{\bf Lattice vibrations can carry angular momentum and magnetic moments under broken inversion or time-reversal symmetry, forming so-called chiral phonons. While such excitations have been explored in nonmagnetic systems via optical probes, their direct detection in magnetic materials and coupling to spin excitations remain largely unexplored. Here, using neutron spectroscopy, sensitive to both nuclear and magnetic scattering, we reveal the magnetic signature of chiral phonons in ferrimagnetic \fzmo with Curie temperature $T_{\rm C}\sim49$~K. Below $T_{\rm C}$, we observe enhanced magnetic scattering of phonons at small momenta, arising from strong magnon-phonon coupling. In addition, out-of-plane intensity modulation, phonon mode splitting, and field-induced Zeeman shifts are observed, all closely associated with the ferrimagnetic order. These features vanish above $T_{\rm C}$, where phonon spectra are dominated by nuclear scattering. These observations demonstrate the existence of chiral phonons carrying substantial magnetic moments that directly contribute to magnetic scattering, and establish neutron spectroscopy as a powerful, momentum-resolved probe of their magnetic character.}
\end{abstract}
\maketitle

Chiral phonons are collective vibrational modes in crystals, characterized by circularly polarized ionic motions carrying nonzero angular momentum. They were initially proposed and observed in two-dimensional systems with broken inversion symmetry\cite{PhysRevLett.112.085503,PhysRevLett.115.115502,Chen_2019,doi:10.1126/science.aar2711}, where the rotational mode propagates within the plane of rotation. Subsequent studies extended these ideas to three-dimensional chiral crystals, including $\alpha$-HgS\cite{Ishito2023}, quartz\cite{Ueda2023} and Te\cite{Zhang2023}, where chiral phonons were demonstrated to exhibit nonzero helicity\cite{PhysRevLett.127.125901,zhang2025}, with their angular momentum having a finite projection along the phonon propagation direction---a feature absent in their two-dimensional counterparts. The chirality and angular momentum of phonons have emerged as fundamental concepts in lattice dynamics, giving rise to a variety of exotic phenomena,  including Einstein-de Haas effect\cite{PhysRevLett.112.085503,Dornes2019,ZhangH2025}, phonon magnetism\cite{Nova2017,doi:10.1126/science.adi9601,Basini2024,Davies2024}, phonon Zeeman effect\cite{PhysRevMaterials.1.014401,PhysRevMaterials.3.064405,PhysRevB.105.094305}, and phonon Hall effect\cite{PhysRevLett.95.155901,PhysRevLett.96.155901,PhysRevLett.105.225901,Grissonnanche2020}. More recently, increasing attention has focused on the interplay between chiral phonons and other degrees of freedom, including charge, spin, orbital, and topology\cite{PhysRevResearch.4.013129,PhysRevLett.127.186403,PhysRevB.110.094401,PhysRevB.110.024423,PhysRevLett.133.246604,Zhang2024,Cheng2020,PhysRevLett.128.075901,doi:10.1126/sciadv.adj4074,Cui2023,doi:10.1073/pnas.2304360121,Bao2023,Wu2023,PhysRevLett.134.196906,PhysRevLett.134.196905}, opening new avenues for manipulating phonon-driven phenomena in complex quantum materials.

Despite these advances, most existing studies have focused on nonmagnetic systems\cite{PhysRevLett.112.085503,PhysRevLett.115.115502,Chen_2019,doi:10.1126/science.aar2711,Ishito2023,Ueda2023,Zhang2023,PhysRevLett.127.125901,zhang2025,PhysRevResearch.4.013129,PhysRevLett.127.186403,PhysRevB.110.094401,Cheng2020,PhysRevLett.128.075901,doi:10.1126/sciadv.adj4074}, leaving the coupling between chiral phonons and magnetic order, particularly their interactions with magnons, largely unexplored \cite{PhysRevB.110.024423,PhysRevLett.133.246604,Cui2023,doi:10.1073/pnas.2304360121,Wu2023,Bao2023,PhysRevLett.134.196906,PhysRevLett.134.196905}. Experimentally, investigations have so far relied primarily on optical techniques, such as Raman scattering\cite{Ishito2023,Zhang2023,PhysRevLett.134.196906,Cui2023,Wu2023}, infrared and terahertz spectroscopy\cite{doi:10.1126/science.aar2711,Cheng2020,PhysRevLett.128.075901,doi:10.1126/sciadv.adj4074}, and resonant inelastic X-ray scattering\cite{Ueda2023,ueda2025}. These methods utilize the (pseudo-)angular momentum transfer between circularly polarized photons and chiral phonons\cite{PhysRevLett.115.115502,Ishito2023,Zhang2023,zhang2025}, but are inherently limited in their ability to map the full energy-momentum dispersions of chiral phonons.

\begin{figure}[htb]
	\centering
	\includegraphics[width=0.95\linewidth]{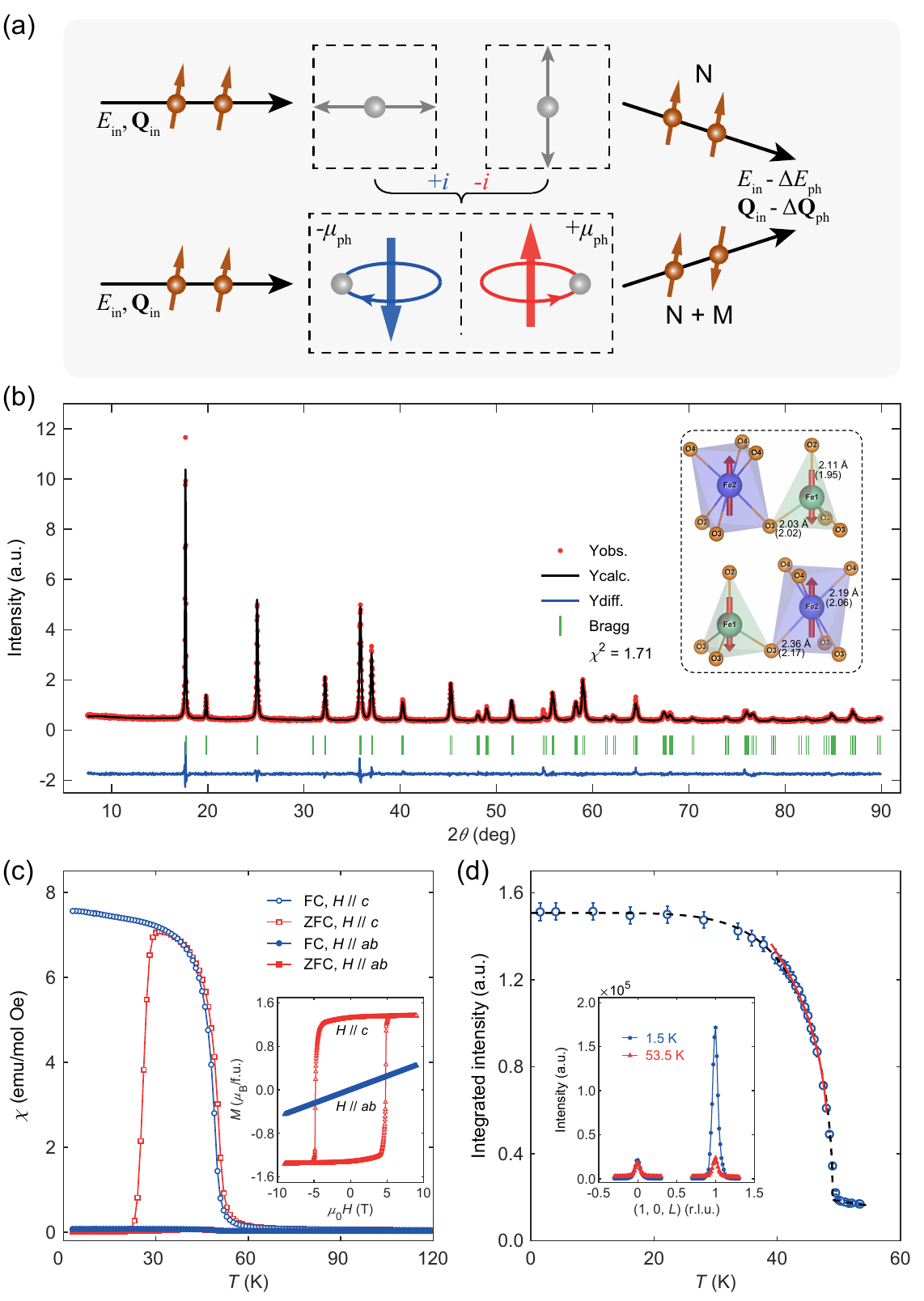}
	\caption{\label{fig:characterizations}{(a) Schematic of neutron probing of lattice vibrations: conventional linear phonons give only nuclear (N) scattering, whereas chiral phonons with phonon magnetic moments ($\mu_{\rm ph}$) yield both N and magnetic (M) signals. (b) Room-temperature powder XRD pattern and Rietveld refinement. Inset: ferrimagnetic structure and Zn-induced lattice distortions in FZMO, with Fe--O bond lengths labeled (values for FMO in parentheses). (c) Magnetic susceptibility of an FZMO single crystal under zero-field-cooled (ZFC) and field-cooled (FC) conditions in a 0.1-T field. Inset: $M$-$H$ curves at 10~K. (d) Temperature dependence of the integrated intensity of the (1,\,0,\,1) magnetic Bragg peak measured on EIGER. Error bars denote one standard deviation throughout the paper. The red curve is a power-law fit and the dashed line is a guide to the eye. Inset: elastic scans along (1,\,0,\,$L$) at 1.5 and 53.5~K.}}
\end{figure}

In contrast, neutron scattering offers a powerful alternative by providing simultaneous access to both nuclear and magnetic scattering cross sections of chiral phonons, thereby enabling momentum-resolved studies. As illustrated in Fig.~\ref{fig:characterizations}(a), in conventional linear phonon scattering processes, the signal arises predominantly from nuclear scattering, reflecting the displacement of atomic nuclei. In principle, neutrons can also probe the magnetic response associated with phonon excitations, which in the context of chiral phonons originates from their intrinsic phonon magnetic moments---a consequence of the circular motion of charged ions. These moments are typically on the order of the nuclear magneton ($\mu_{\rm N}$)\cite{PhysRevMaterials.1.014401,PhysRevMaterials.3.064405,PhysRevB.105.094305}, far smaller than electronic spin or orbital moments measured in Bohr magnetons ($\mu_{\rm B}$), and thus usually produce negligible magnetic scattering. However, in systems with strong electron-, orbital-, or spin-phonon coupling\cite{Cheng2020,PhysRevLett.128.075901,doi:10.1126/sciadv.adj4074,Cui2023,doi:10.1073/pnas.2304360121,Wu2023,Bao2023}, the effective magnetic moment of chiral phonons can be significantly enhanced, reaching the $\mu_{\rm B}$ scale. Under such conditions, magnetic scattering from chiral phonons becomes appreciable, opening a pathway for momentum-resolved investigations of chiral phonons and their coupling to various static orders and collective excitations\cite{PhysRevResearch.4.013129,PhysRevLett.127.186403,PhysRevB.110.094401,PhysRevB.110.024423,PhysRevLett.133.246604,Zhang2024,Cheng2020,PhysRevLett.128.075901,doi:10.1126/sciadv.adj4074,Cui2023,doi:10.1073/pnas.2304360121,Wu2023,Bao2023,PhysRevLett.134.196906,PhysRevLett.134.196905}.

Fe$_{2-x}$Zn$_x$Mo$_3$O$_8$ is a type-I multiferroic system exhibiting colossal linear magnetoelectricity and a giant thermal Hall effect\cite{PhysRevX.5.031034,Wang2015,Ideue2017,PhysRevLett.131.136701}, indicative of strong spin-lattice coupling. In previous studies of the parent compound \fmo (FMO), we observed topological magnon polarons arising from strong magnon-phonon coupling\cite{Bao2023} and a pair of degenerate chiral phonon modes carrying substantial magnetic moments, up to 0.68~$\mu_{\rm B}$\cite{Wu2023}. The magnetic configuration of FMO can also be tuned by external fields or chemical doping, inducing a metamagnetic transition from an antiferromagnetic to a ferrimagnetic state\cite{PhysRevX.5.031034,Wang2015}. These properties establish FMO and its derivatives as an ideal platform to study how magnetic-order symmetry influences chiral phonons under strong magnon-phonon coupling, while the sizable phonon magnetic moments facilitate direct, momentum-resolved detection by neutron scattering.

In this work, we perform inelastic neutron scattering (INS) to investigate the excitation spectra of 12.5\% Zn-doped \fzmo (FZMO), which hosts a collinear ferrimagnetic ground state below $T_{\rm C}\sim$ 49~K. At 6~K, the excitation spectra reveal three intense magnon branches, along with signatures of hybridization and gap openings between magnon and phonon bands, indicative of strong magnon-phonon coupling. Low-energy phonons generate measurable magnetic scattering, particularly at small momenta. These modes exhibit out-of-plane intensity modulation, pronounced splitting, and field-induced Zeeman shifts, all closely linked to the underlying ferrimagnetic order. These features vanish above $T_{\rm C}$. Our results provide a rare momentum-resolved mapping of chiral phonons and their magnetic signatures in magnetic systems.

\begin{figure*}[htb]
	\centering
	\includegraphics[width=0.90\linewidth]{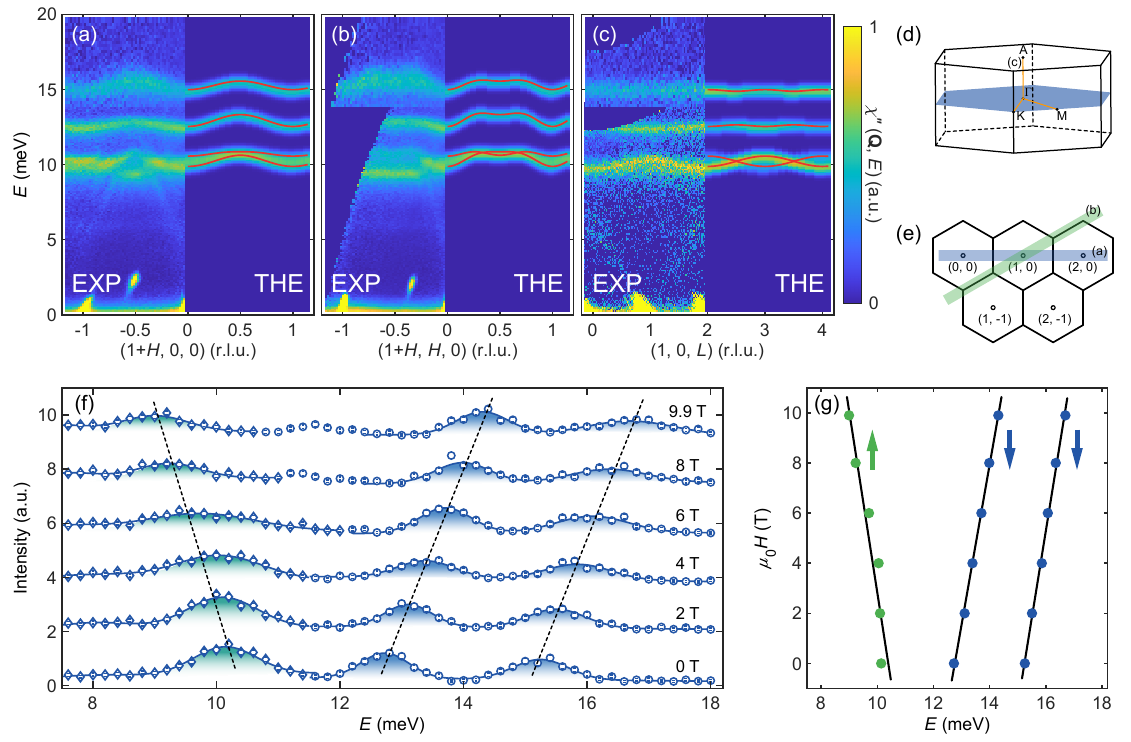}
	\caption{\label{fig:magnons}{(a)-(c) INS spectra of FZMO at 6~K along the [100], [110], and [001] (left), compared with calculated spin-wave spectra (right). Data were measured on 4SEASONS with $E_{\rm i}$ = 18 and 30~meV, scaled to incoherent elastic scattering. Integration ranges are listed in Table~S3\cite{sm}. (d),\,(e) Brillouin zone in 3D and its 2D projection with high-symmetry paths. (f) Energy scans at (1.2,\,0,\,0) under varying magnetic fields, measured on EIGER. Diamonds and circles denote data collected in normal- and high-resolution configurations, respectively, with vertical offsets applied for clarity. Curves are Gaussian fits and dashed lines are guides to the eye. (g) Extracted peak positions as a function of field, compared with model predictions (solid lines). Up and down arrows denote spin quantum numbers $\Delta S_z=+1$ and $-1$.}}
\end{figure*}

\begin{figure*}[htb]
	\centering
	\includegraphics[width=0.95\linewidth]{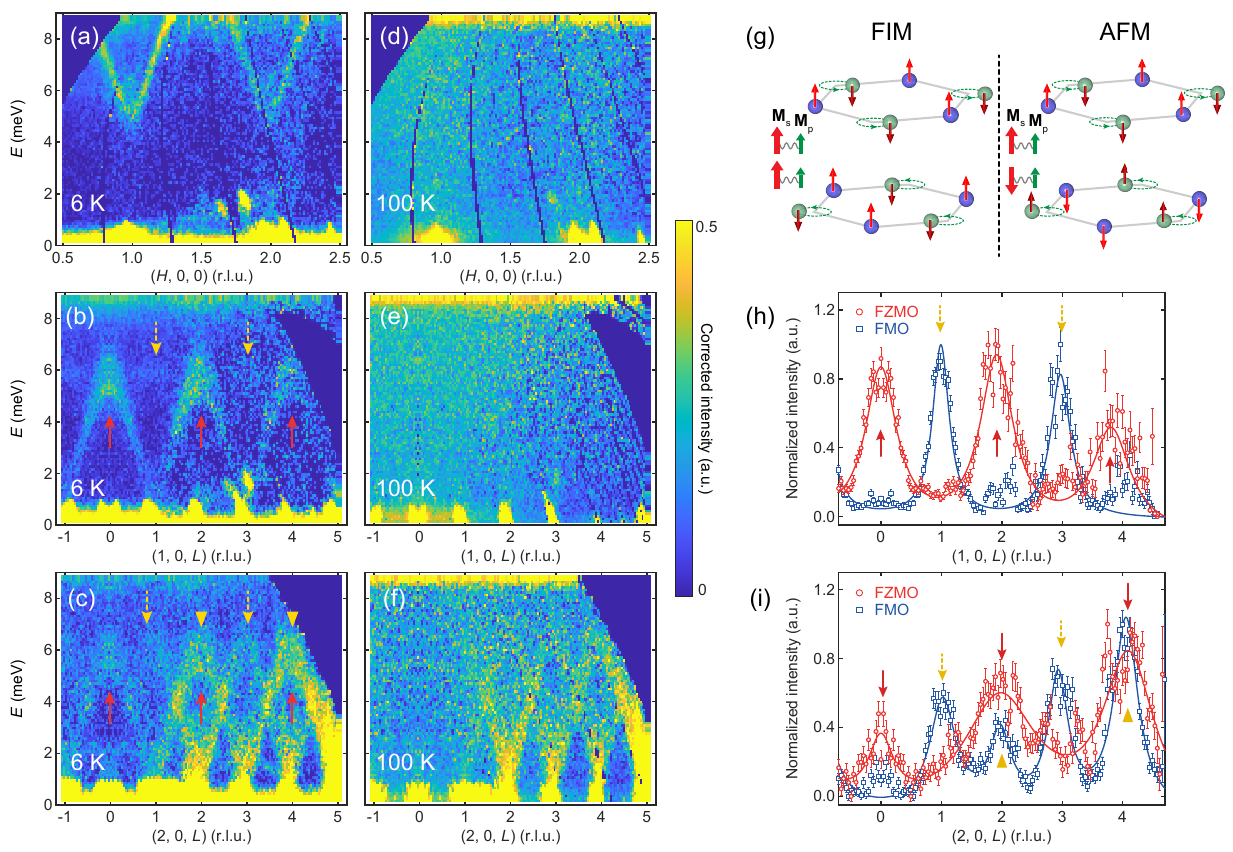}
	\caption{\label{fig:phonons}{(a)-(c) Phonon spectra at 6~K along ($H$,\,0,\,0), (1,\,0,\,$L$), and (2,\,0,\,$L$). (d)-(f) Corresponding spectra at 100~K. Data with $E_{\rm i} = 12$~meV are corrected by Bose factor. (g) Schematic illustrating the coupling between phonon magnetic moments ($M_{\rm p}$) arising from in-plane rotation of Fe1 ions (right-handed example shown) and net spin moments ($M_{\rm s}$) in adjacent layers for ferrimagnetic (FIM) and antiferromagnetic (AFM) states. (h) Constant-$E$ cuts along (1,\,0,\,$L$) for FZMO (red circles) and FMO (blue squares). (i) Corresponding cuts along (2,\,0,\,$L$). The energy interval is $5.4\pm0.4$~meV. Curves are Lorentzian fits. Solid (dashed) arrows mark intense zone-center chiral optical phonons in FZMO (FMO), while triangles indicate dominant nuclear scattering contributions in (h), (i); the same symbols appear in (b), (c) for guidance.}}
\end{figure*}

\begin{figure}[htb]
	\centering
	\includegraphics[width=0.95\linewidth]{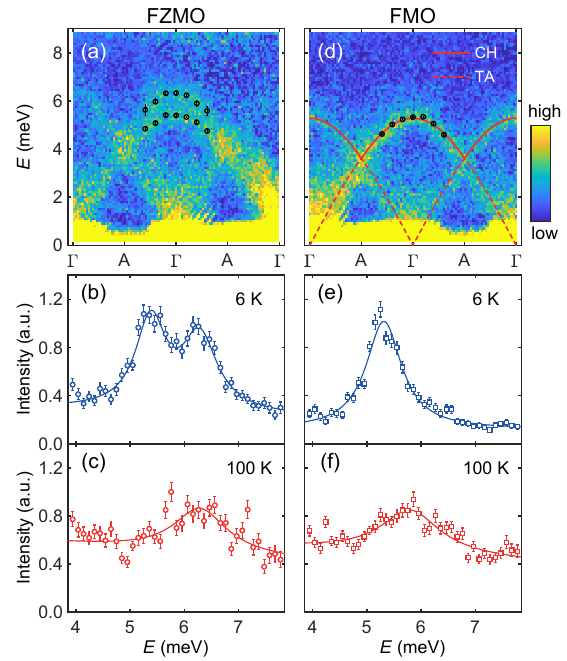}
	\caption{\label{fig:comparison}{(a) Phonon spectra of FZMO at 6~K along [001], integrated over equivalent Brillouin zones. Black circles mark peak positions from Lorentzian fits to various constant-$q$ cuts. (b),\,(c) Constant-$q$ cuts at zone center for 6~K and 100~K. (d)-(f) Corresponding data for FMO. Dashed and solid lines denote transverse acoustic (TA) and chiral optical phonon (CH) modes, respectively.}}
\end{figure}

Single crystals of FZMO were grown using chemical vapor transport method (see Note 1 in Supplemental Material (SM) for details \cite{sm}). Powder X-ray diffraction (XRD) shows that Zn preferentially occupies the tetrahedral Fe1 sites while retaining the polar $P6_{3}mc$ symmetry, consistent with earlier studies\cite{varret1972etude,bertrand1975etude,PhysRevX.5.031034}. Rietveld refinements indicate that Zn substitution primarily perturbs the coordinating oxygen positions and induces local lattice distortions, particularly around the octahedral Fe2 environment (inset of Fig.~\ref{fig:characterizations}(b) and Note 2 in SM\cite{sm}), likely modifying the crystal field and single-ion anisotropy. Magnetic susceptibility measurements reveal ferrimagnetic ordering below $T_{\rm C}\sim$49~K with strong easy-axis anisotropy, in agreement with elastic neutron results [Figs.~\ref{fig:characterizations}(c) and~\ref{fig:characterizations}(d)]. This ferrimagnetic state results from selective dilution of the Fe1 sublattice, altering the balance of interlayer couplings (Note 3 in SM\cite{sm}). A power-law fit yields a critical exponent $\beta=0.170$, slightly above the 2D Ising value observed in FMO\cite{Bao2023,collins1989magnetic}.

Figures~\ref{fig:magnons}(a)-\ref{fig:magnons}(c) show the overall excitation spectra of FZMO measured along high-symmetry directions [Figs.~\ref{fig:magnons}(d) and~\ref{fig:magnons}(e)]. Guided by our previous study on FMO\cite{Bao2023}, we identify three types of excitations within the same energy-momentum window: (i) three intense, weakly dispersive magnon branches at higher energies (10--20~meV); (ii) several weak, strongly dispersive phonon modes originating from low energies; and (iii) magnon-phonon hybrid excitations---magnon-polarons---emerging near the intersections between the magnon and phonon bands.

Notably, the three intense magnon branches exceed the two nearly doubly-degenerate branches expected from linear spin-wave theory for a bipartite ferrimagnet\cite{PhysRevB.111.024407}. This anomaly is captured by a minimal spin model incorporating site-dependent anisotropy and interlayer coupling (Note 3 in SM\cite{sm}). The model yields two nearly degenerate Fe1-derived modes with $\Delta S_z = +1$, and two separate Fe2-derived modes with $\Delta S_z = -1$. The latter separation is attributed to Zn-substitution-induced local distortions, which creates two distinct Fe2 environments with different single-ion anisotropies, consistent with structural considerations and prior reports of stronger anisotropy at Fe2 sites in the parent compound (Notes 2 and 3 in SM\cite{sm}). The field evolution further supports this assignment: the lowest branch softens while the two higher branches harden with increasing field [Figs.~\ref{fig:magnons}(f) and~\ref{fig:magnons}(g)], consistent with their opposite $\Delta S_z$ signs.

Having identified the distinct excitation types, we now examine the low-energy phonon spectra to understand how strong magnon-phonon coupling shapes the lattice dynamics in the ferrimagnetic state. At 6~K, phonon excitations are visible across both small and large momenta [Figs.~\ref{fig:phonons}(a)-\ref{fig:phonons}(c)]. In contrast, at 100~K, their intensity is strongly suppressed at small momenta [Figs.~\ref{fig:phonons}(d) and~\ref{fig:phonons}(e)], but remains visible at larger momenta [Fig.~\ref{fig:phonons}(f)]. The enhanced spectral weight at small momenta at 6~K arises from additional magnetic scattering due to off-resonant magnon-phonon coupling, whose magnitude depends on the coupling strength, the detuning between the two excitations, and the magnon magnetic moment\cite{Wu2023,Wu2025}. This magnetic contribution is superimposed on the nuclear phonon signal. At 100 K, it disappears as the ferrimagnetic order vanishes, and the phonon intensities follow the expected momentum dependence of purely nuclear scattering, becoming more prominent at larger momenta.

A detailed examination of the phonon spectra reveals a saddle point near 5.5~meV, featuring a minimum along [100] and a maximum along [001] shown in Figs.~\ref{fig:phonons}(a)-\ref{fig:phonons}(c). Our previous studies have shown that such zone-center excitations correspond to a pair of degenerate chiral phonons that carry substantial phonon magnetic moments in antiferromagnetic FMO\cite{Wu2023,Bao2023}. In ferrimagnetic FZMO, similar excitations persist but exhibit a distinct out-of-plane intensity modulation. Specifically, enhanced phonon intensity appears at even $L$, marked by solid arrows in Figs.~\ref{fig:phonons}(b) and~\ref{fig:phonons}(c), whereas FMO shows odd-$L$ modulation, indicated by dashed arrows. A detailed comparison is provided by constant-$E$ cuts across the saddle points in Figs.~\ref{fig:phonons}(h) and~\ref{fig:phonons}(i). Exceptions occur at large momenta such as (2,\,0,\,2) and (2,\,0,\,4), where strong nuclear scattering dominates, as also observed at 100~K [Fig.~\ref{fig:phonons}(f)]. It is worth noting that the 12.5\% Zn substitution in FZMO does not alter the parent polar structure [Fig.~\ref{fig:characterizations}(b)], and the nuclear dynamic structure factor alone cannot account for the reversed intensity modulation. This is confirmed by measurements at 100~K, where FZMO and FMO exhibit nearly identical phonon spectra without any $L$-dependent modulation [Figs.~\ref{fig:phonons}(d)-\ref{fig:phonons}(f); Ref.~\onlinecite{Bao2023}]. Instead, the reversal emerges only below the magnetic ordering temperatures, indicating its close connection to the underlying magnetic configurations. A plausible explanation is that the phonon modes acquire a magnetic component, i.e., chiral phonons carrying angular momenta and effective magnetic moments, which couple to the spin order, as illustrated in Fig.~\ref{fig:phonons}(g). The distinct coupling forms, arising from the relative orientations of phonon magnetic moments and net spin moments in adjacent layers, lead to constructive interference at even $L$ and extinction at odd $L$ in the ferromagnetic stacking of FZMO, whereas the opposite pattern appears in the antiferromagnetic stacking of FMO. A detailed analysis is provided in Note 4 of the SM\cite{sm}.

In addition to the spectral weight modulation, the ferrimagnetic order also induces an intrinsic splitting of the chiral phonons. As shown in Fig.~\ref{fig:comparison}(a), the phonon spectra along the [001] direction exhibit a clear splitting within the optical branches at 6~K. A constant-$q$ cut at the zone center shows a pronounced splitting of approximately 1~meV, corresponding to $\sim$20\% of the phonon energy [Fig.~\ref{fig:comparison}(b)]. At 100~K, this splitting vanishes, and the phonon modes recover their original profiles [Fig.~\ref{fig:comparison}(c)]. In contrast, no appreciable splitting is observed in FMO [Figs.~\ref{fig:comparison}(d)-\ref{fig:comparison}(f)], confirming that this effect is closely tied to the ferrimagnetic order in FZMO.

It is worth noting that the split phonon modes along the [001] direction correspond to truly chiral phonons with nonzero helicity, allowed by the discrete threefold rotational symmetry of the crystal\cite{Ishito2023,Ueda2023,Zhang2023,PhysRevLett.127.125901,zhang2025}. In both the antiferromagnetic and paramagnetic phases [Figs.~\ref{fig:comparison}(c)-\ref{fig:comparison}(f)], these excitations arise from the doubly degenerate $E_2$ modes, primarily involving in-plane motion of Fe1 ions\cite{Wu2023}. Upon entering the ferrimagnetic phase [Figs.~\ref{fig:comparison}(a) and \ref{fig:comparison}(b)], time-reversal symmetry breaking lifts this degeneracy, yielding two chiral phonon branches with opposite helicities. The splitting appears with the onset of ferrimagnetic order\cite{Wu2025}. Our zone-center constant-$q$ cuts under magnetic field reveal opposite field dependences for the split doublet, consistent with polarization-resolved magneto-Raman measurements\cite{Wu2025}, providing additional evidence for their chiral nature (Note 5 in SM\cite{sm}). This mechanism contrasts with earlier reports in which phonon splittings originated from inversion symmetry breaking\cite{doi:10.1126/science.aar2711,Ishito2023,Ueda2023,Zhang2023,PhysRevLett.128.075901,doi:10.1126/sciadv.adj4074,ueda2025}. Moreover, the observed splitting and population imbalance between phonons of opposite chirality are consistent with the enhanced phonon thermal Hall conductivity in the ferrimagnetic phase of Fe$_{2-x}$Zn$_x$Mo$_3$O$_8$\cite{Ideue2017}.

A similar magnetic-order-induced splitting of chiral phonons has been reported in the ferromagnetic Weyl semimetal Co$_3$Sn$_2$S$_2$\cite{PhysRevLett.134.196906,PhysRevLett.134.196905}. However, there the chiral phonons primarily originate from nonmagnetic S atoms\cite{PhysRevLett.134.196906,PhysRevLett.134.196905}, and the large energy separation between phonons and magnons renders magnon-phonon coupling negligible. In contrast, FZMO hosts strong magnon-phonon coupling, producing distinct spectral signatures in the chiral phonon branches (Fig.~\ref{fig:phonons}). Recently, the concept of axial phonons has been introduced as a broad class of phonons that carry angular momentum and an associated magnetic moment, while chiral phonons form a symmetry-defined subset that additionally breaks improper rotational symmetry\cite{Juraschek2025}. This framework naturally accounts for (i) the absence of splitting and the vanishing chiral term for the degenerate optical phonons in FMO, and (ii) the transverse acoustic phonons in both compounds (Fig.~\ref{fig:comparison}), which also exhibit magnetic signatures arising from axial phonon magnetic moments without requiring the symmetry constraints of true chirality.

We note that a recent theoretical study proposed that INS can distinguish chiral from linear phonons via angle-resolved measurements and by tracking phonon splitting under applied magnetic fields\cite{wang2025INS}. This approach considers the distinct contributions of rotational atomic motion to the nuclear dynamic structure factor. However, the magnetic scattering of chiral phonons arising from phonon magnetic moments remains largely unexplored. Polarized INS may thus provide a promising route to directly resolve phonon chirality in momentum space. Unlike magnons, whose chirality arises from the circular precession of ordered magnetic moments\cite{PhysRevLett.125.027201,PhysRevB.108.104404}, phonon chirality originates from circular ionic motion producing effective magnetic moments. Polarized neutrons probe these phonon magnetic moments rather than the atomic rotation itself, making the detection of phonon chirality distinct from that of magnons. Developing the corresponding theoretical framework and experimental implementation represents an important direction for future research.

In summary, we present the first complete mapping of truly chiral phonon dispersions across the Brillouin zone and reveal their magnetic signature in ferrimagnetic FZMO. Below $T_{\rm C}$, we observe enhanced magnetic scattering, out-of-plane intensity modulation, intrinsic mode splitting, and field-induced Zeeman shifts, all consistent with chiral phonons carrying substantial magnetic moments. These findings establish a direct link between magnetic order, magnon excitations, and chiral phonons in systems with strong magnon-phonon coupling, and demonstrate that neutron scattering provides a powerful, momentum-resolved probe of chiral phonons via their inherent phonon magnetic moments.

\bigskip

$Acknowledgments$---We thank Yuan Wan, Fangliang Wu, Qi Zhang and Zhao-Long Gu for stimulating discussions. The work was supported by the National Natural Science Foundation of China (Grant No.~12225407), the National Key Projects for Research and Development of China (Grants No.~2021YFA1400400 and No.~2024YFA1409200), the National Natural Science Foundation of China (Grants No.~12434005, No.~12404173, and No.~12204160), the Natural Science Foundation of Jiangsu Province (Grants No.~BK20233001, No.~BK20241251, and No.~BK20241250), the Natural Science Foundation of the Higher Education Institutions of Jiangsu Province (Grant No.~23KJB140012), the China Postdoctoral Science Foundation (Grants No.~BX20240161 and No.~2024M751367), the Jiangsu Funding Program for Excellent Postdoctoral Talent (Grant No.~2024ZB021), the Xiaomi Young Scholars - Technology Innovation Award, and the Fundamental Research Funds for the Central Universities (Grants No.~KG202501 and No.~14380251). Part of this work is based on experiments performed at the Swiss spallation neutron source SINQ, Paul Scherrer Institute, Villigen, Switzerland. This work is also supported by J-PARC (Proposal No.~2020B0002) and ISIS (Proposal No.~2410072, DOI: 10.5286/ISIS.E.RB2410072).

\bigskip

$Data~availability$---The data are not publicly available. The data are available from the authors upon reasonable request.


%

\end{document}